**Metal-Organic Chemical Vapor Deposition of PtSe$_2$**

Maximilian Prechtl[1], Marc Busch[1], Oliver Hartwig[1], Kangho Lee[1], Tanja Stimpel-Lindner[1], Cormac Ó Coileáin[1], Kuanysh Zhussupbekov[2], Ainur Zhussupbekova[2], Samuel Berman[2], Igor V. Shvets[2], and Georg S. Duesberg*[1]

[1] Institute of Physics, Faculty of Electrical Engineering and Information Technology and SENS Research Centre, University of the Bundeswehr Munich, 85577 Neubiberg, Germany

[2] CRANN, School of Physics, Trinity College Dublin, Dublin 2, Ireland

*Corresponding author e-mail: duesberg@unibw.de

**Abstract**

Platinum diselenide (PtSe$_2$), a novel two-dimensional material from the class of noble-metal dichalcogenide (NMD), has recently received significant attention due to its outstanding properties. PtSe$_2$, which undergoes a semi metallic to semiconductor transition when thinned, offers a band-gap in the infrared range and good air stability. These properties make it a prime active material in optoelectronic and chemical sensing devices. However, a synthesis method that can produce large-scale and reliable high quality PtSe$_2$ is highly sought after. Here, we present PtSe$_2$ growth by metal organic chemical vapor deposition. Films were grown on a variety of centimeter scale substrates and were characterized by Raman, X-ray photoelectron and X-ray diffraction spectroscopy, as well as scanning tunneling microscopy and spectroscopy. Domains within the films are found to be up to several hundred nanometers in size, and atomic scale measurements show their highly ordered crystalline structure. The thickness of homogenous films can be controlled via the growth time. This work provides fundamental guidance for the synthesis and implementation of high quality, large-scale PtSe$_2$ layers, hence offering the key requirement for the implementation of PtSe$_2$ in future electronic devices.

# 1. Introduction

Two dimensional (2D) materials are the rising star of current materials research. Two element compound 2D materials with the general formula $MX_2$ (M = Metal, X = chalcogen) – the transition metal dichalcogenides (TMDs) – began to raise increasing interest as they offer unique electronic[1,2], photonic[3], optoelectronic[4] and sensing properties[5,6]. One of the most recently research material class within the field of 2D materials are the noble-metal dichalcogenides. For those materials, the metal (M) is typically either palladium or platinum. NMDs like $PdSe_2$ or $PtSe_2$ offer, in addition to their unique properties[7–11], high-stability in air[12,13] which is a key requirement for device integration.

One special material within the NMD family is $PtSe_2$ offering in total six possible crystallographic phases[14] with the 1T phase being the most commonly experimentally observed. Notably, it undergoes a semimetal to semiconductor transition when thinned down[15,16] along with a band gap in the infrared (IR) range[7,17] and carrier mobilities[12] which is among the top performers when compared to other TMD materials. $PtSe_2$ has shown to be a prime material for chemical sensing devices[6,18], optoelectronic IR waveguide detectors[19] but has also proven its viability when used as transistor[15] or in thermoelectric devices due to its high Seebeck coefficient[20]. Nevertheless, a high-quality, large-scale and reproducible synthesis approach of $PtSe_2$ is still highly sought after, potentially exploiting the materials unique properties in modern devices.

To date, a few synthesis techniques for thin $PtSe_2$ layers have been reported, which are mainly chemical vapor deposition[21–24] (CVD) and thermal assisted conversion (TAC)[6,12,18,19]. In TAC pre-deposited Pt layers are converted with Se vapor to $PtSe_2$ which allows growth of continuous films with controlled thickness on the large scale. TAC synthesis can be performed at low temperatures allowing even synthesis on flexible substrates[25]. Conformal and selective deposition has been reported with TAC in combination with atomic layer deposition of the Pt film. However, TAC grown films[6] suffer from a low crystallinity with the individual $PtSe_2$ domains in the films having typically sizes of a few nanometers[18,25,26] which sets a fundamental limit to the film quality, and, furthermore it is challenging to yield monolayer films. CVD growth of $PtSe_2$ yields monolayer individual flakes , in particular when using crystalline substrates such as sapphire substrates [21–24], however, it remains challenging to grow continuous films at larger scales. Hence, an improved growth technique is required, that combines the layer quality of CVD with the versatility of TAC synthesis.

In this work, we present for the first time the metal-organic (MO)CVD growth of $PtSe_2$ which is a highly reproducible synthesis route combining a MO-Pt-precursor source and $H_2Se$ to scalable grow high-quality $PtSe_2$ layers on a variety of commonly used substrates such $SiO_2$, c-plane sapphire and pyrolytic carbon (PyC). A custom-built synthesis setup fully encapsulated in a glovebox was used for the $PtSe_2$ growth, allowing versatility for deposition and the growth on areas up to 6'' wafers. The quality of the grown films is evaluated using Raman spectroscopy, X-ray photoelectron spectroscopy (XPS), atomic force microscopy (AFM), X-Ray diffraction spectroscopy (XRD) which together demonstrate high material quality. Furthermore, the atomic and electronic structure of the deposited layers is investigated by scanning tunneling microscopy (STM) and spectroscopy (STS), which reveal domain sizes within the layers of up to 300 nm. Within those domains, a defect free growth was visible in STM measurements. This study should give a growth and integration route of high-quality, large-scale PtSe$_2$ layers, laying the cornerstone for integration of PtSe$_2$ into modern devices.

## 2. Results and Discussion

### 2.1: Synthesis of large-scale PtSe$_2$ on various substrates

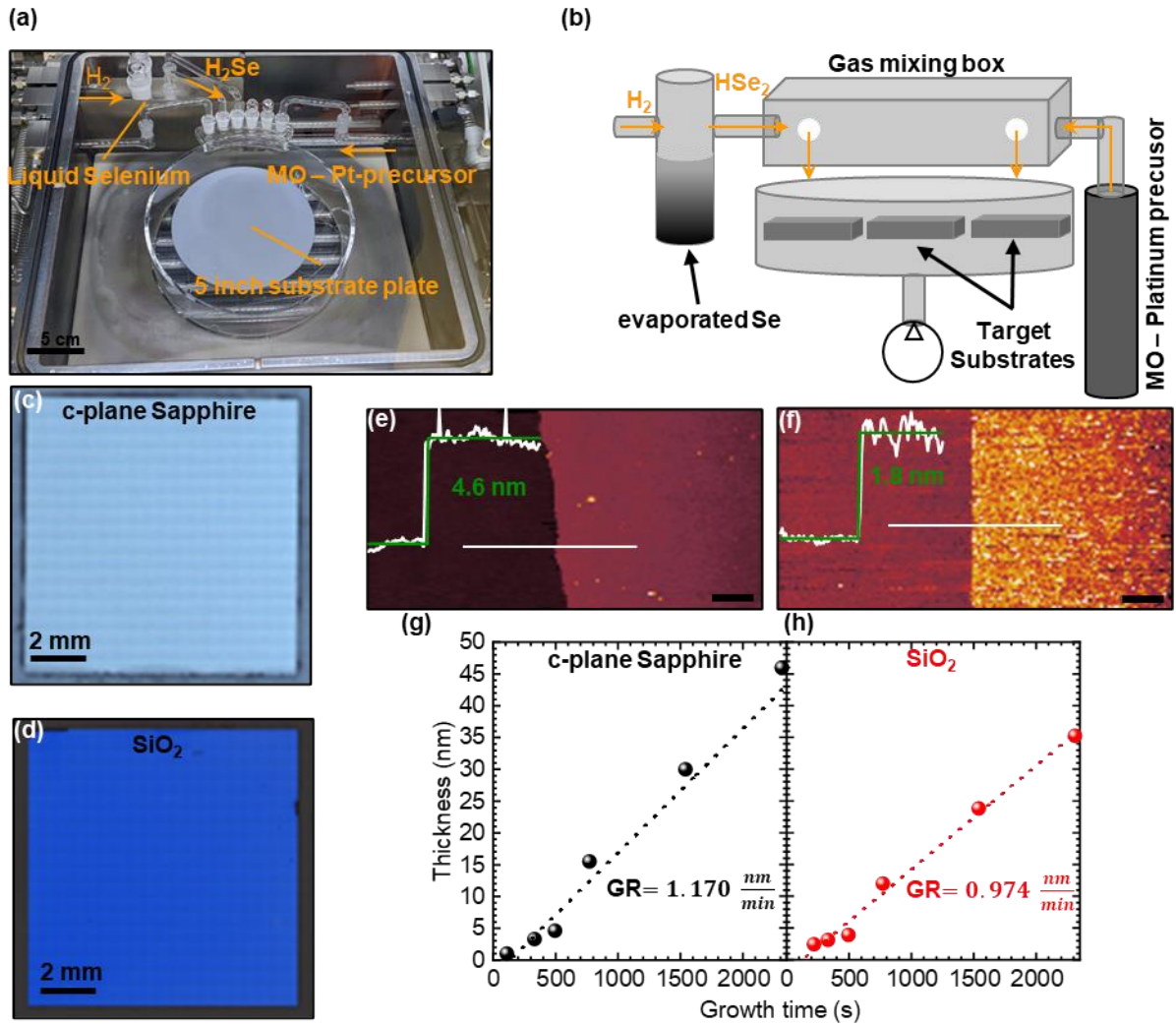

**Figure 1: Synthesis of PtSe$_2$ by MOCVD.** (a) Photograph of the 6'' custom built MOCVD reactor. (b) Schematic of the reactor chamber (c) 1 nm and (d) 3.2 nm PtSe$_2$ layer homogenously deposited on 1 * 1 cm$^2$ scale (c) c-plane sapphire and (d) SiO$_2$ substrates. (e) AFM image including height profile (white) and fit (green) for a 4.6 nm thick PtSe$_2$ grown on c-plane sapphire and (f) 1.8 nm PtSe$_2$ grown on SiO$_2$, profile positions are marked as white lines. The scale bars (black) are 1 μm. (g, h) PtSe$_2$ layer thickness versus growth time on c-plane sapphire and SiO$_2$ respectively. Growth rates have been calculated from linear fitting.

Large-scale PtSe$_2$ films have been synthesized by MOCVD in a custom-built cold-wall reactor shown in Figure 1 (a). A schematic of the growth chamber, showing gas flows during deposition, is shown in Figure 1 (b). The substrates are placed on a 6'' substrate plate. H$_2$Se is produced in-situ from evaporated selenium powder heated to 225 °C in a compartment off the primary reactor and flushed with H$_2$. (Trimethyl)ethylcyclopentadienylplatinum(IV) ((CH$_3$)$_3$(C$_2$H$_2$C$_5$H$_4$)Pt is the platinum precursor provided in a Swagelok© cylinder, which is externally heated to 40 °C and supplied through heated pipes. Prior to deposition the substrates were annealed in a H$_2$ atmosphere for 15 min at 600 °C. The growth of the PtSe$_2$ is typically caried out at 600 °C. Homogenous deposition was achieved on 1 * 1 cm$^2$ c-plane sapphire (Figure 1 (c)) and SiO$_2$ (Figure 1 (d)) substrates. Additional SEM images of the deposited layers can be found in the Supplementary information.

The growth method allows a high degree of control over the thickness by adjusting the growth duration. The growth rates (GRs) on different substrates were determined by AFM measurements (Figure **1** (e, f)) after deposition and result in $GR(sapphire) = 1.170 \frac{nm}{min}$ for c-plane sapphire (Figure 1 (g)) substrates and $GR(SiO_2) = 0.974 \frac{nm}{min}$ for $SiO_2$ substrates (Figure 1 (h)). Notably, the growth rate is significantly lower on $SiO_2$ compared to c-plane sapphire even though the intercepts with the x-axis for the linear fits, which define when successful seeding on the target substrate is achieved, are at 129 s and 120 s for c-plane sapphire and $SiO_2$, respectively. However, no deposition could be observed on $SiO_2$ for a deposition time of 110 s by Raman and AFM measurements, a growth time which had already led to a 1 nm thick film on c-plane sapphire. From this, one can conclude that a closed layer is formed significantly faster on the c-plane sapphire (once seeding is achieved) than on the $SiO_2$ substrates.

**2.2 Layer quality analysis by Raman spectroscopy, XPS and XRD measurements**

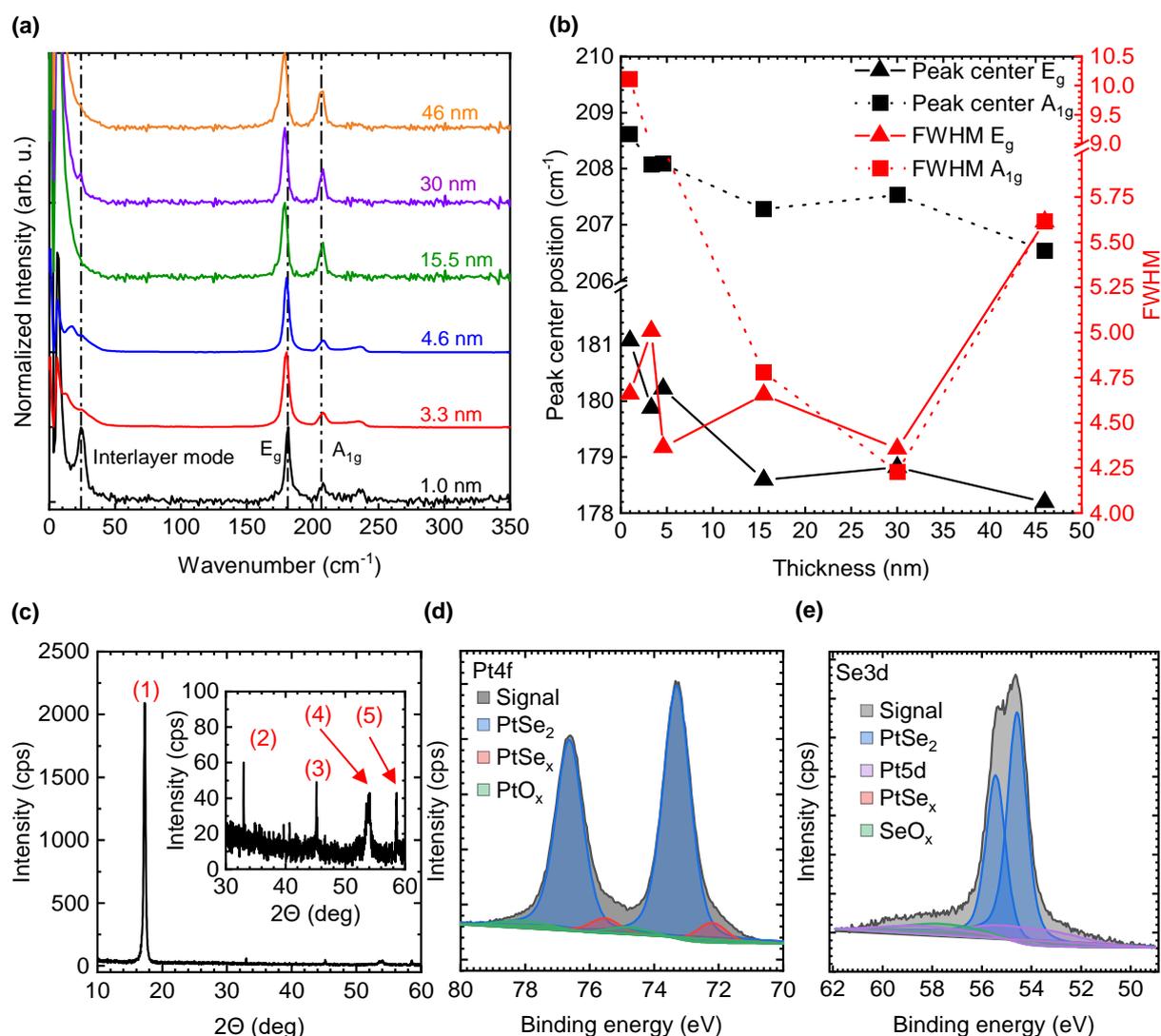

**Figure 2: Raman spectra, XRD and XPS analysis of MOCVD grown PtSe₂.** (a) Normalized, averaged Raman spectra of differently thick $PtSe_2$ films grown on c-plane sapphire. (b) Peak center of the $E_g$ (black triangle) and $A_{1g}$ (black square) mode and respective FWHM (red) for different thickness $PtSe_2$ films. (c) XRD spectrum of a 35 nm thick $PtSe_2$ film. Inset showing a magnification of the same plot. XPS of the Pt4f core level region (d) and Se3d core level region (e). The spectra have been fitted to highlight the individual contributions to the raw signal (black).

To investigate the material quality, Raman spectra were acquired for PtSe$_2$ films of different thicknesses grown on c-plane sapphire substrates. The spectra were averaged over a total of 100 measured, distributed over the whole sample to increase statistical significance (Figure 2 (a)). A detailed Raman analysis for PtSe$_2$ grown on SiO$_2$ substrates can be found in the supplementary information. For all spectra, the fundamental PtSe$_2$ fingerprint E$_g$ and A$_{1g}$ modes can be identified, while also a prominent interlayer mode located at 24 cm$^{-1}$ is present in the spectrum of the 1 nm PtSe$_2$ (black). The absence of the interlayer mode for thicker layers is expected as it becomes less prominent for increasing film thickness[27]. The evolution of the peak center position and FWHM for the E$_g$ and A$_{1g}$ mode are shown in (Figure 2 (b)). For the E$_g$ (A$_{1g}$) mode a red shift from 181.07 cm$^{-1}$ (208.62 cm$^{-1}$) to 178.2 cm$^{-1}$ (206.54 cm$^{-1}$) is observed which is in good agreement with literature[27–29]. The FWHM of the E$_g$ mode, which is directly correlated with layer quality[29,30], ranges from 4.3 cm$^{-1}$ to 5.6 cm$^{-1}$, giving an indication of high layer quality[29]. The XRD spectrum (Figure 2 (c)) shows a strong feature at 17° which can be attributed to the (001) peak of PtSe$_2$[30,31], the higher order features (inset) at 33° and 45° are attributed to the (101) and (102) peaks, respectively, whose clear presence indicate the high crystallinity of the PtSe$_2$ film. For chemical binding analysis of the synthesized PtSe$_2$, spectra of the core level binding energy of the Pt4f (Figure 2 (d)) and Se3d (Figure 2 (e)) regions are acquired by XPS measurements. The core level region of the C1s peak (not shown here) was also recorded as reference to account for any residual surface charging not compensated by the tool built-in neutralization. The raw signal acquired from the Pt4f peak (grey) region consists of a total of three peaks. The main contribution (blue) has its maximum at 73.3 eV, indicating that 90 % of the Pt-atoms are present in the form of PtSe$_2$. About 6.8 % (red) of the contributing Pt-atoms are present in a sub-stochiometric form PtSe$_x$. The remaining 3.2 % (green) are found in a slightly higher binding state which is attributed to surface oxidation. The latter two contributions are mainly present on the surface of the films, where a loss of selenium atoms (PtSe$_x$), and consequently oxidation (PtO$_x$) is expected[32]. The raw signal of the Se3d core level region (Figure 2 (e), grey) consists of three doublets and a singlet. The main contribution with 88.2 % (blue) originates from PtSe$_2$, while 1.7 % and 10.1 % are attributed to PtSe$_x$ and SeO$_x$, respectively. The last contribution (purple) stems from the Pt5p orbital which was not taken into calculation for the PtSe$_2$ ratio. Furthermore, the survey spectrum (Supplementary Information) did not show any significant presence of other chemical components on our sample. In summary, detailed Raman analysis and XRD measurements confirm the high-quality growth of PtSe$_2$ by our newly developed MOCVD approach. In addition, XPS measurements reveal a stoichiometric ratio of 1:1.83, and that the signals of the Pt4f and Se3d peaks nearly exclusively originate from PtSe$_2$, supporting the conclusions drawn from Raman and XRD analysis.

**2.3 Structural and electronic investigation by atomically resolved STM and STS measurements**

STM (Figure 3 (a, b)) and STS (Figure 3 (c - i)) were used to investigate the quality of PtSe$_2$ films deposited on PyC. Individual PtSe$_2$ domains with sizes of 50 nm to 300 nm can be identified in large-scale STM images, as shown in Figure 3 (a). The high quality of the PtSe$_2$ within such domains is recognizable at smaller scales (Figure 3 (b)). For further evaluation of the atomic structure within one domain, STS measurements (Figure 3 (c - i)) were acquired from the atomically clean region marked with a black square in Figure 3 (b). In Figure 3 (c) I/V (red) and dI/dV (black) spectra can be seen. From the dI/dV curve (black), the semi-metallic character of the 35 nm thick PtSe$_2$ film is confirmed (Figure 3 (c)). Furthermore, atomically resolved images acquired from bias voltages ranging from -0.4 V (Figure 3 (d)) to 0.4 V (Figure 3 (i)) show a perfectly ordered PtSe$_2$ lattice without any surface defects, once again underlining the high quality of PtSe$_2$ produced by our newly developed MOCVD approach.

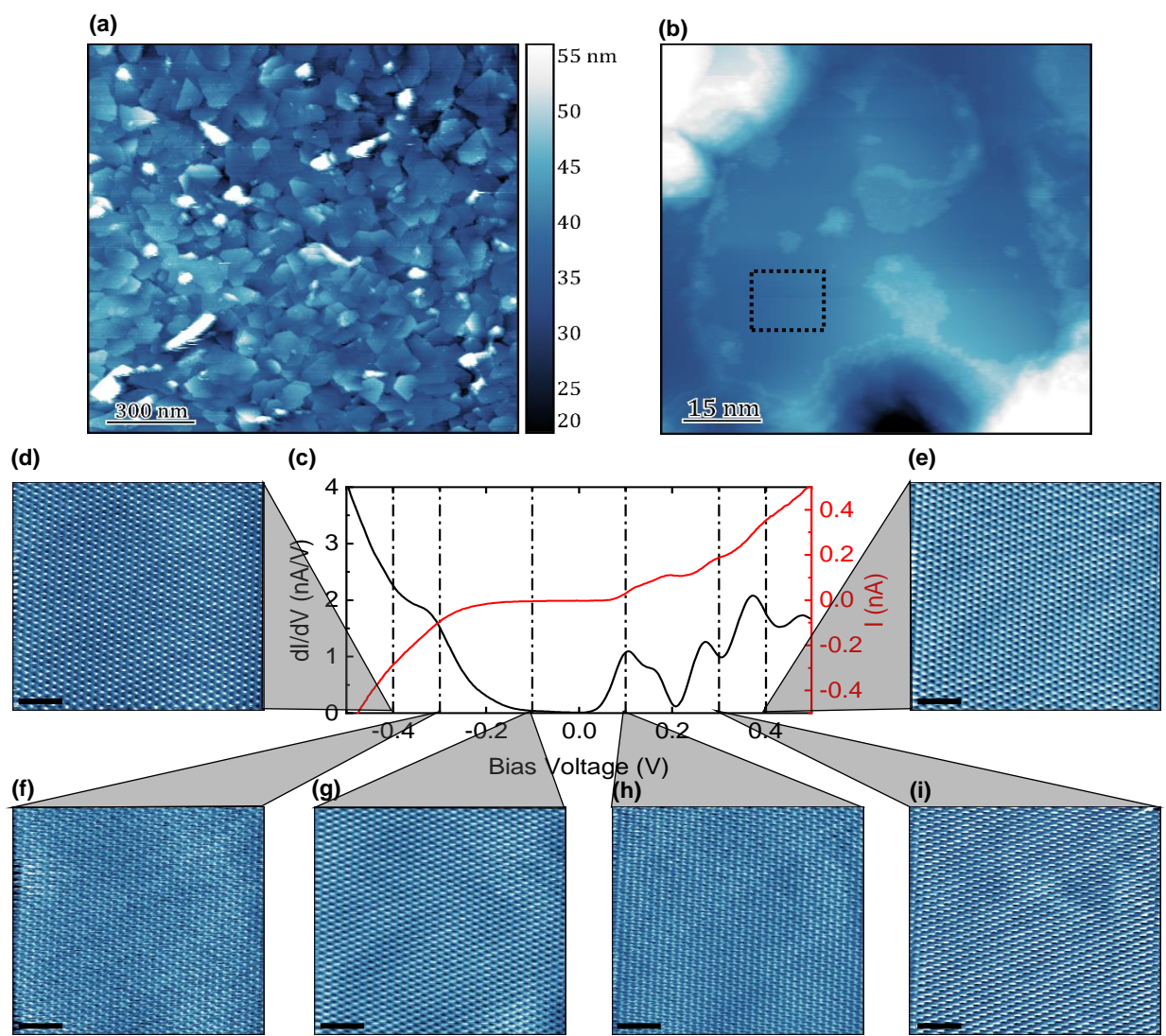

**Figure 3: STM & STS of a 35 nm thick PtSe$_2$ film on PyC.** (a, b) Large-scale STM image of PtSe$_2$ grown on PyC. Domain sizes in the range of 50 nm to 300 nm are visible. The black square in (b) indicated the area where high resolution STM and STS measurements have been conducted. (c) Derivation of current (black) and current (red) for different bias voltages used in STS. (d-i) Atomically resolved structure of PtSe$_2$ obtained for different bias voltages. The scale bars are 2 nm, the total size of all scans is 13 x 13 nm$^2$.

## 3. Conclusion

PtSe$_2$ was grown by combining a MO platinum precursor source and in-situ prepared H$_2$Se. Centimeter-scale synthesis was shown on SiO$_2$, c-plane sapphire and PyC substrates. The high quality of the synthesized material was verified by Raman, XPS and XRD measurements. Atomically resolved STM measurements revealed domains up to 300 nm wide and hexagonal PtSe$_2$ single crystals. Furthermore, atomically resolved STM images have confirmed locally defect free ordering within the single crystals, underlining the high material quality. In addition, the deposition was shown to be successful in a 6'' wafer tool, paving the way towards industry scale synthesis of PtSe$_2$ and its future implementation into real world devices.

**4. Experimental Section:**

*AFM analysis:* A Jupiter XR AFM in tapping mode was used for acquisition of AFM data within this study. OMCL-AC160TS tips were used in tapping mode, the tapping frequency was set in between 200 to 400 kHz. Processing of the acquired raw data and extraction of topographical information was done using Gwyddion (64bit) 2.60.

*Characterization by Raman and X-ray photoelectron spectroscopy:* All Raman data was acquired using a WITec Alpha 300 Confocal Raman system. For all measurements, a green laser (532 °C) with a power of 2 mW (on sample surface) was used, with a 100x objective (NA = 0.9). A grating with 1800 lines per mm was used. All Raman data shown are averaged data from 100 points gather from a 1 x 1 cm² $PtSe_2$ film. Data was analyzed and processed using software provided with the tool. For XPS data a Versa Probe XPS tool (Physical Electronics GmbH) using the monochromated Al Kα line (1486.7 eV). Single spectra were acquired using 61.8 W beam power paired with a 100 µm spot size. Spectra have been fitted with Gaussian-Lorentzian fit functions using Multi-Pak.

*X-ray diffraction spectroscopy:* High resolution XRD analysis was obtained using Bruker D8 Discover equipment with a monochromated Cu K-alpha source.

*Scanning tunneling microscopy and spectroscopy:* STM and STS measurements were conducted on a low temperature microscopy (Createc) at liquid nitrogen temperature (77 K). The base pressure in the main STM chamber was $5*10^{-11}$ mbar. Data were obtained in constant current mode using a single-crystalline W(001) tip.

**Supporting Information:**

The data that supports the findings of this study are available from the corresponding author upon reasonable request.

**Acknowledgements:**

This work was financially supported by the European Commission under the projects ULISSES [825272], QUEFORMAL [829035] and Graphene Flagship [881603], as well as the German Ministry of Education and Research (BMBF) under the projects NobleNEMS [16ES1121] and ACDC [13N15100]. Further we thank the UniBW M and the DTEC project Vital Sense for support. We would like to thank the REK Innovation GmbH, namely Dr. Venzi Rangelov, for their in-depth support on the technical side of our deposition tool K.Z. and A.Z. would like to acknowledge funding from IRC through GOIPD/2022/774 and GOIPD/2022/443 awards.

**Author contributions**

M.P., C.C. and G.S.D. conceived and defined the study. G.S.D and I.S. supervised the study. M.P. and M.B. fabricated and conducted Raman and AFM measurements. A.Z. conducted XRD measurements and analyzed the data. O.H. conducted XPS measurement. T.S.L. and O.H. analyzed the spectra by multiple peak fitting. K.Z. and S.B. acquired STM and STS images, C.C., K.Z. and I.S. analyzed the respective data. K.L. acquired and analyzed electrical data. All authors extensively discussed and reviewed the manuscript. M.P., C.C. and G.S.D. wrote the manuscript with inputs from all authors.

**Competing interests:**

The authors declare no competing interests.

**Metal-Organic Chemical Vapor Deposition of PtSe$_2$**

Maximilian Prechtl[1], Marc Busch[1], Oliver Hartwig[1], Kangho Lee[1], Tanja Stimpel-Lindner[1], Cormac Ó Coileáin[1], Kuanysh Zhussupbekov[2], Ainur Zhussupbekova[2], Samuel Berman[2], Igor V. Shvets[2], and Georg S. Duesberg*[1]

## Supporting Information

[1] Institute of Physics, Faculty of Electrical Engineering and Information Technology and SENS Research Centre, University of the Bundeswehr Munich, 85577 Neubiberg, Germany

[2] CRANN, School of Physics, Trinity College Dublin, Dublin 2, Ireland

*Corresponding author e-mail: duesberg@unibw.de

# 1. SEM images of PtSe$_2$ grown on SiO$_2$

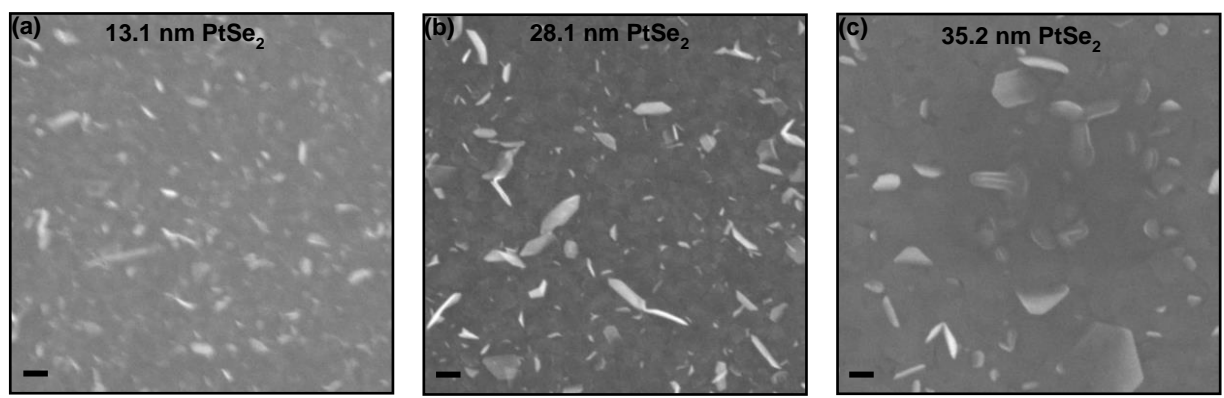

**Figure S1: SEM imaging of PtSe$_2$ grown on different substrates.** SEM images of MOCVD grown PtSe$_2$ on (a – c) SiO$_2$, (d – f) c-plane sapphire and (g – l) pyrolytic carbon. All scale bars are 100 nm.

**Figure S1** shows SEM images for 13.1 nm thick (a), 28.1 nm thick (b) and 35.2 nm thick (c) PtSe$_2$ layers synthesized on SiO$_2$ substrates. The hexagonal shape of the deposited flakes are clearly visible.

## 2. Raman Analysis of MOCVD grown $PtSe_2$ on $SiO_2$

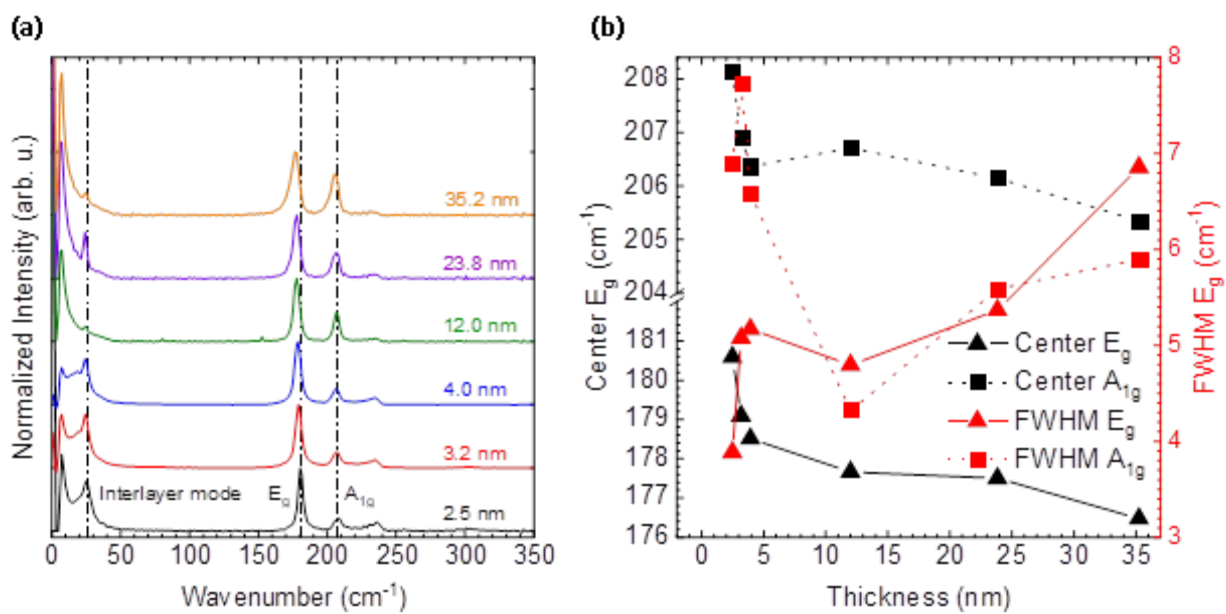

**Figure S2: Raman analysis of MOCVD grown $PtSe_2$ on $SiO_2$ substrates.** (a) Normalized, averaged Raman spectra of different thickness $PtSe_2$ films grown on c-plane sapphire. (b) Peak center of the $E_g$ (black triangle) and $A_{1g}$ (black square) mode and respective FWHM (red) for different thickness $PtSe_2$ films.

Figure S2 (a) shows six Raman spectra acquired from different thicknesses of $PtSe_2$ deposited on $SiO_2$ using the MOCVD growth approach presented in the main text. The spectra have been averaged (100 spots equally distributed on a 1 x 1 $cm^2$ sample) and normalized to the $E_g$ mode for better comparison. For all thicknesses, the Raman fingerprint mode of $PtSe_2$ ($E_g$ and $A_{1g}$) can be identified. With the peak position of the $E_g$ ($A_{1g}$) mode located at 180.6 $cm^{-1}$ (208.12 $cm^{-1}$) for the 2.5 nm thick $PtSe_2$ (black), a blue shift of 3.9 $cm^{-1}$ (3.7 $cm^{-1}$) can be identified with increasing film thickness up to 35.2 nm (Figure S2 (b)). This is in good agreement with literature[1–3]. Furthermore, a prominent interlayer mode evolving around 24 $cm^{-1}$ can be identified for the thinnest layer, becoming less prominent with increasing film thicknesses[4]. A FWHM of 3.9 $cm^{-1}$ is extracted from the $E_g$ mode of the thinnest layer, increasing to 6.86 $cm^{-1}$ for the thickest $PtSe_2$ layer. This indicated high layer quality, as the FWHM of the $E_g$ mode has been reported to be a quality indicator[2].

## 3. XPS survey spectra of MOCVD synthesized PtSe$_2$

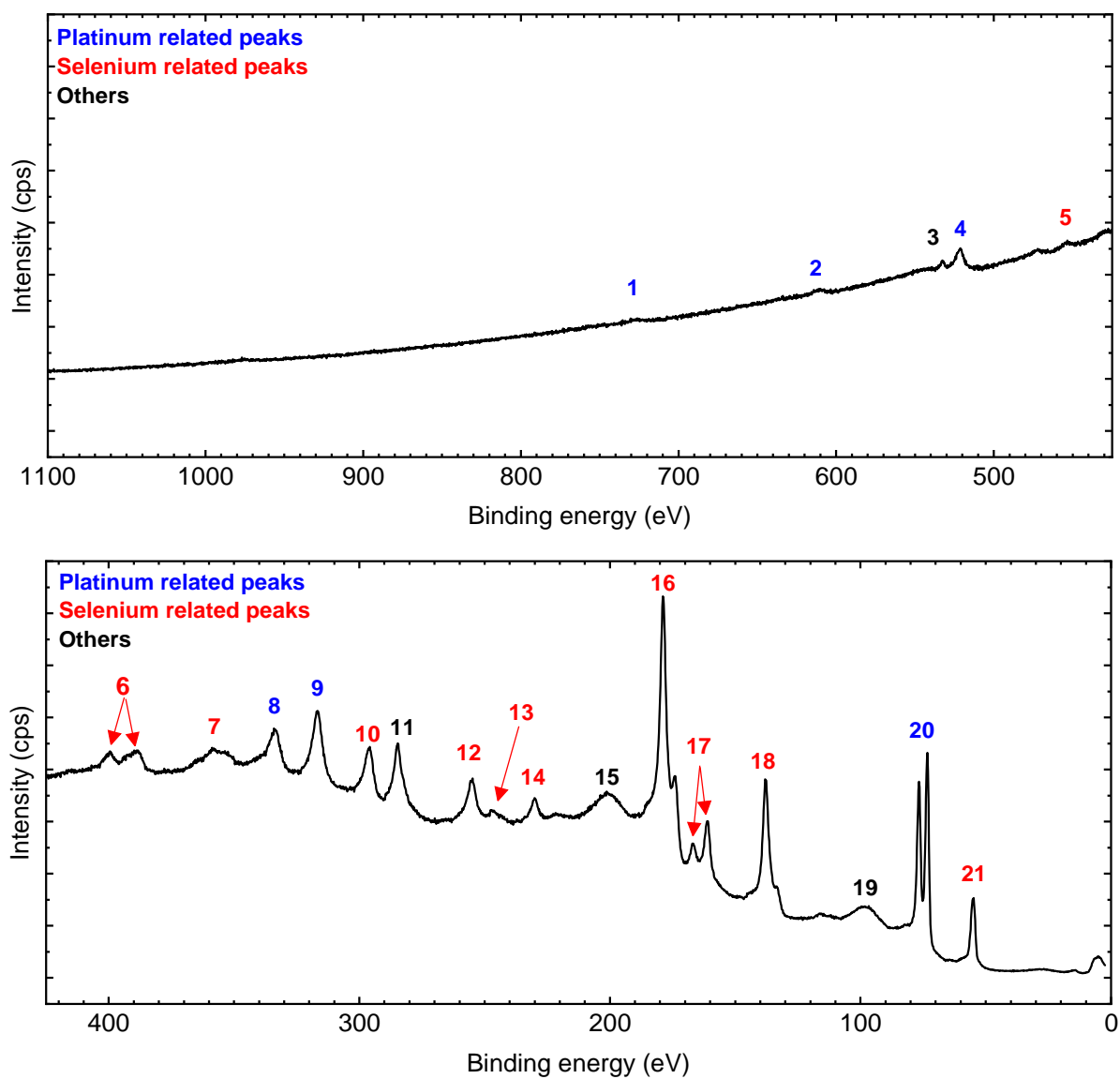

**Figure S3: XPS survey spectrum.** XPS survey spectrum of MOCVD 13 nm thick PtSe$_2$ deposited on SiO$_2$ for binding energies (a) 1100 eV to 425 eV and (b) 425 eV to 0 eV. The primary platinum related peaks have been marked in blue, selenium related peaks have been marked in red and the remaining peaks have been left as black.

For investigation of possible contaminants in our MOCVD grown PtSe$_2$ we acquired a high-resolution survey spectrum (Figure S3) of a 13 nm thick PtSe$_2$ sample deposited on SiO$_2$. The spectrum has been shifted using the C1s peak (284.6 eV[5]) as a reference. All peaks in the spectra have been labelled while platinum and selenium marked features are marked in blue and red, respectively[5]. Other features, originating from plasmon losses, carbon and oxygen have been marked in black. A conclusive summary of all features is given in Table S1. Notably, a strong nonlinear background was observed during acquisition. The background originates from plasmon losses which are caused by the sample surface morphology[6]. This is well supported by strong plasmon peak features (label 15 and 19). In addition to selenium and platinum related features, we can identify a small amount of oxygen on the sample (label 3) which we attribute to surface oxidation occurring on the PtSe$_2$ layer. Nevertheless, with the exception of oxygen, the overall signal originates from selenium and platinum related features, showing a high purity of the layers without any surface contaminations.

**Table S1: Summary table of all peaks which are visible in Figure S3.**

| Label in figure | Peak origin | Peak position(measured) [eV] | Literature value[5] [eV] |
|---|---|---|---|
| 1 | Pt4s | 725 | 725 |
| 2 | Pt4p$_{1/2}$ | 609 | 609 |
| 3 | O1s | 532 | 531 |
| 4 | Pt4p$_{3/2}$ | 521 | 520 |
| 5 | Se | 453 | - |
| 6 | Se | 399 & 389 | - |
| 7 | Se | 359 | - |
| 8 | Pt4d$_{3/2}$ | 334 | 332 |
| 9 | Pt4d$_{5/2}$ | 316 | 315 |
| 10 | Se Auger line L$_3$M$_{23}$M$_{45}$ ($^1$P) | 297 | 299 |
| 11 | Carbon 1s | 284.6 (shift reference) | 298.4 |
| 12 | Se Auger line L$_2$M$_{23}$M$_{45}$ ($^1$P) | 255 | 257 |
| 13 | Se | 248 | - |
| 14 | Se3s | 231 | 232 |
| 15 | Energy loss peak of 16 | | |
| 16 | Se Auger line L$_3$M$_{45}$M$_{45}$ ($^1$P) | 179 | 181 |
| 17 | Se3p$_{1/2}$ & Se3p$_{3/2}$ | 167 & 161 | 169 & 163 |
| 18 | Se Auger line L$_2$M$_{45}$M$_{45}$ | 138 | 140 |
| 19 | Energy-loss peak of 20 | 99 | - |
| 20 | Pt4f$_{5/2}$ & Pt4f$_{7/2}$ | 77 & 73 | 74 & 71 |
| 21 | Se3d$_{3/2}$ & Se3d$_{5/2}$ & Pt5p | 55 | 74 & 71 |
| 21 | Se3d$_{3/2}$ & Se3d$_{5/2}$ & Pt5p | 55 | 57 & 56 & 52 |